\documentclass[sigconf]{acmart}
\AtBeginDocument{%
  \providecommand\BibTeX{{%
    \normalfont B\kern-0.5em{\scshape i\kern-0.25em b}\kern-0.8em\TeX}}}

\usepackage{caption}
\usepackage{romannum}
\usepackage{enumitem}
\usepackage{setspace}
\usepackage{multicol}
\usepackage{pifont}
\usepackage{multirow}
\usepackage{graphicx}
\usepackage{colortbl}
\newcommand{\circled}[1]{\ding{\numexpr171+#1\relax}}
\usepackage{pdftexcmds}
\usepackage[T1]{fontenc}
\RequirePackage{fontawesome}
\usepackage{hyperref}

\copyrightyear{2024}
\acmYear{2024}
\setcopyright{rightsretained}
\acmConference[WWW '24 Companion]{Companion Proceedings of the ACM Web Conference 2024}{May 13--17, 2024}{Singapore, Singapore} \acmBooktitle{Companion Proceedings of the ACM Web Conference 2024 (WWW '24 Companion), May 13--17, 2024, Singapore, Singapore}\acmDOI{10.1145/3589335.3651464}
\acmISBN{979-8-4007-0172-6/24/05}
\makeatletter
\gdef\@copyrightpermission{
  \begin{minipage}{0.3\columnwidth}
   \href{https://creativecommons.org/licenses/by/4.0/}{\includegraphics[width=0.90\textwidth]{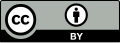}}
  \end{minipage}\hfill
  \begin{minipage}{0.7\columnwidth}
   \href{https://creativecommons.org/licenses/by/4.0/}{This work is licensed under a Creative Commons Attribution International 4.0 License.}
  \end{minipage}
  \vspace{5pt}
}
\makeatother

\begin{document}

\title[Exploring Academic References on Stack Overflow]{A Tale of Two Communities: \\
Exploring Academic References on Stack Overflow}


\author{Run Huang}
\orcid{0009-0009-7467-4896}
\affiliation{%
  \institution{University of Southern California}
  \city{Los Angeles}
  \country{USA}
  }
\email{runhuang@usc.edu}

\author{Souti Chattopadhyay}
\orcid{0000-0003-1644-7344}
\affiliation{%
  \institution{University of Southern California}
  \city{Los Angeles}
  \country{USA}
  }
\email{schattop@usc.edu}


\begin{abstract}
Stack Overflow is widely recognized by software practitioners as the go-to resource for addressing technical issues and sharing practical solutions. While not typically seen as a scholarly forum, users on Stack Overflow commonly refer to academic sources in their discussions. Yet, little is known about these referenced academic works and how they intersect the needs and interests of the Stack Overflow community. To bridge this gap, we conducted an exploratory large-scale study on the landscape of academic references in Stack Overflow. Our findings reveal that Stack Overflow communities with different domains of interest engage with academic literature at varying frequencies and speeds. The contradicting patterns suggest that some disciplines may have diverged in their interests and development trajectories from the corresponding practitioner community. Finally, we discuss the potential of Stack Overflow in gauging the real-world relevance of academic research.

% Stack Overflow is a popular platform among practitioners to ask questions about practical challenges and receive rapid technical solutions. Users engaging in Stack Overflow discussions often refer to academic sources, with over 12,000 posts referring to more than 10,000 academic articles. However, we lack an understanding of how research from the academic community intersects the needs and interests of the Stack Overflow community. In this study, we analyze the patterns and networks of academic references in Stack Overflow. Our findings reveal that Stack Overflow communities with different domains of interest engage with academic literature at varying frequency and speed. The contradicting patterns suggest that some disciplines diverge in their interests and trajectories from the corresponding practitioner community. Finally, we discuss the potential of Stack Overflow in evaluating the practical applicability of academic research.

\end{abstract}





\maketitle

\section{Introduction}
% Over the past decades, we've seen rapid advancements in the field of Computer Science, accompanied by an exponential surge in research papers. Yet, how many of them have been integrated into the industry and contribute to technological inventions and products? Are these research providing value to software developers and more generally the computer science community? 

% What is the problem?
Stack Overflow (SO) is a popular Q\&A platform among software practitioners and developers~\cite{sodoccit} for discussing technical issues~\cite{Zhou2019BountiesOT}. These discussions might go beyond just finding quick technical fixes, as SO users often delve deeper into the problems at hand and share their own knowledge and expertise in the process~\cite{Zhang2021ReadingAO}. Organically, SO users refer to a wide range of sources to support their claims~\cite{repeatlinks}, including academic articles published in conferences and journals. Such practices suggest that SO might be acting as a potential bridge between the two communities---developers and academic researchers---for knowledge dissemination.

% SO users engage in discussions to identify feasible solution pathways, build broader insights about the problem, or look for examples of implementation~\cite{overflow2017stack}. Organically, SO users often refer to various sources to support their discussions in various capacities~\cite{repeatedlysharedsolink}. Many of these references lead to articles from the academic community published in conferences and journals. This indicates that SO provides a bridge between the two communities---practitioners and academic researchers---to share knowledge.

However, we currently lack an understanding of these referenced articles and how the SO community engages with them. This limits our insights into SO's role in facilitating knowledge transfer and reflecting the practical relevance of academic research. Investigating this gap could uncover potential discrepancies between cutting-edge research and prevailing industry norms and interests. It could also guide both communities in identifying how academic advancements can match up the needs of the SO community and the broader developer community it serves. To the best of our knowledge, this is the first study to examine the recognition of academic research within SO, and to investigate how scientific knowledge is consumed or utilized for solving real-world software development challenges across different communities of interest on SO.

% However, we lack an understanding of how the SO community interacts with academic sources and how academic knowledge contributes to SO discussions. This hinders us from understanding SO's role in the knowledge transfer process and further assessing the practical relevance of academic research. Investigating this gap will can help us identify the misalignment between cutting-edge research and industry practices and interests. It will further guide both communities in identifying how academic advancements can match the needs of the SO or even the developer community at large. To the best of our knowledge, this is the first work to investigate how academic research is recognized and referenced by SO communities and delve into how practitioners can utilize and integrate academic knowledge through SO.

Prior works have studied academic references on Twitter~\cite{tweets} and Wikipedia~\cite{mapofsciencewikipedia}, emphasizing the general visibility and popularity of scientific findings rather than their practical relevance. Following this trend, Altmetric.com~\cite{altmetric} included SO citations as one of its alternative metrics for measuring academic impact. However, it only quantifies how frequently academic articles are mentioned but overlooks the context of these references and their connections to the original discourse. As a result, Altmetric suffers from a similar limitation of serving merely as an indicator of visibility.

In this paper, we explored the landscape of academic references on SO and took an initial step toward assessing their potential to reflect the practical relevance and utility of academic research. We sought to answer the following research questions (\textbf{RQ}s):

\begin{spacing}{0.8}
\begin{enumerate}[itemsep=1.1mm,topsep=1.72mm]
    \item[\textbf{RQ1:}] What academic articles are cited on Stack Overflow?
    \item[\textbf{RQ2:}] Which parts of Stack Overflow rely on academic sources?
    \item[\textbf{RQ3:}] How do various Stack Overflow communities interact with academic research?
    \item[\textbf{RQ4:}] How quickly are academic articles recognized on Stack Overflow?
\end{enumerate}
\end{spacing}

We sifted through 44 million URLs on SO and identified 15,009 references to 10718 unique academic articles. Leveraging topic modeling and social network analysis techniques, we examined the patterns of interaction between SO communities with diverse technical interests and academic research from various fields. To support future research, we have made our dataset publicly available~\cite{soref}.

\section{Methodology}
Unlike Wikipedia or academic settings where citations follow a standard format, users on Stack Overflow (SO) commonly use hyperlinks to reference academic sources. This practice makes it challenging to distinguish academic references from other types of links, as both appear as bare URLs without bibliographic information~\cite{sodoccit}. In this section, we describe our heuristic method for identifying links leading to academic articles and the data collection process.

\textit{\romannum{1}) \underline{Filtering for Potential Candidates}}.\hspace{0.25em} Instead of examining every external link on SO, we narrowed our search to those that are most likely to host academic documents. Prior efforts~\cite{altmetric} have limited to links containing recognized identifiers such as DOIs and ISBNs. We significantly expanded this scope by incorporating links originating from recognized academic repositories, e.g., the \textit{\small ACM Digital Library}.

% containing a Digital Object Identifier (DOI), or originating from recognized academic repositories, such as \textit{\small ACM Digital Library}, since these links are most likely to host academic documents.

Links containing DOIs were identified using regex. For the remaining, we checked if their web domains appeared on a list of domain names affiliated with academic repositories. In compiling this list, we considered various academic entities, including publishers {\small (e.g., Springer)}, academic societies {\small (e.g., ACM)} and databases {\small (e.g., ResearchGate)}. However, given the sheer number of these potential sources, it is infeasible to include them all. Instead, we opted for a best-effort approach, focusing on the most significant ones. To better align with Stack Overflow's emphasis on software and computing, we included all 31 publishers indexed by DBLP, which is a major Computer Science bibliography. With inputs from experts and authoritative sources~\cite{h5index}, we further enrich the selection with 13 academic societies renowned for organizing prestigious conferences {\small (e.g., CVPR)} and ten well-known academic databases. One author then iteratively curated all relevant web domains belonging to these entities (e.g., \href{https://aclanthology.org/}{\textit{\small aclanthology.org}} and \href{https://aclweb.org}{\textit{\small aclweb.org}} of ACL).

Although our selection\footnote{Details on the included academic repositories and methodology are available \href{https://github.com/aceatusc/sciso-www}{at \scriptsize \faExternalLink}} is not exhaustive, it effectively covers a substantial part of the academic publishing landscape. The 31 included publishers issue over 16,000 journals in various fields~\cite{journalcounts}, and the 13 academic societies host more than 500 conferences annually, not to mention the extensive reach of the included databases.

% Although our selection was not exhaustive, it effectively covered a substantial part of the academic publishing landscape. 
% , not to mention the extensive reach of the included databases. Therefore, we believe the broad coverage of our selection enabled us to comprehensively identify potential candidates of academic references on Stack Overflow.

\vspace{1pt}
\textit{\romannum{2}) \underline{Retrieving and Validating Bibliographic data}}.\hspace{0.25em} For every candidate link identified previously, we assumed its content to be academic and attempted to retrieve possible bibliographic information. For links directing to PDF files, we extracted titles and DOIs using Grobid~\href{https://github.com/kermitt2/grobid}{\scriptsize \faExternalLink}, a popular tool for parsing scientific documents. If the link leads to a webpage, we assumed it to be the landing page of a research article, and retrieved its potential titles from various HTML tags, such as \texttt{\small <h1>}, \texttt{\small <meta property="og:title">}, \texttt{\small <title>}, \texttt{\small <h2>}, etc. Additionally, we searched for DOIs within the HTML file using regex. For non-functional links, we accessed their archived versions via Internet Archive's Wayback Machine~\href{https://web.archive.org/}{\scriptsize \faExternalLink}.

In cases where the candidate link does not lead to academic content as we assumed, the surmised bibliographic data extracted heuristically from HTML or PDF files would be spurious. With this in mind, we can filter out ineligible candidates by cross-validating their surmised metadata against two major academic databases: Semantic Scholar and OpenAlex~\cite{priem2022openalex}. A matching record in the title or DOI will confirm the academic nature of the candidate link. We then gathered detailed metadata for verified academic references from the two databases, including abstract, venue, citations, etc.

% \textit{\romannum{3}) Validating and Aggregating Metadata.}\hspace{0.4em} In cases where candidate links did not lead to scholarly documents or landing pages as we assumed, the bibliographic data extracted from the HTML or PDF files were invalid. To identify genuine academic references, we cross-validated each candidate's extracted metadata against two major academic databases: Semantic Scholar and OpenAlex~\cite{priem2022openalex}. A matching record in the title and/or DOI confirmed the academic nature of the link. We then gathered detailed metadata for verified academic references from the two databases, including their abstracts, venues, and citation counts.

\vspace{3pt}
\noindent
\textbf{Data Collection.} Based on the official Stack Overflow data dump released on December 8, 2023, we extracted 44 million URLs found in the edit histories of 59 million posts (totaling 160 million revisions). Similar to prior work~\cite{BrokenSOLink}, URLs embedded within code blocks were omitted, as they are often irrelevant to knowledge sharing. After this refinement, we obtained a dataset~\cite{soref} of 30.9 million links, from which 15,009 references to 10718 academic articles made by 12,963 posts were identified through the previously described process.

% talk about DOI only consititudes 19.13% of all identified reference, so Altmetric (the only existing data source) is incomplete

\section{Results}

\noindent
\emph{\textbf{RQ 1}: What academic articles are cited on Stack Overflow?}

For a fine-grained characterization of the cited articles, we categorize each article into its corresponding Field of Research (FoR) using OpenAlex's concept tagging model~\href{https://docs.google.com/document/d/1OgXSLriHO3Ekz0OYoaoP_h0sPcuvV4EqX7VgLLblKe4/edit}{\scriptsize \faExternalLink}. This model analyzes the abstract and title and generates an initial list of relevant research fields, along with a confidence score. The fields are assigned in six levels, from the broadest to highly specific, following Wikidata's taxonomy. For each article, we selected the second level field (e.g., \textit{\small World Wide Web}) with the highest confidence score as its FoR.

The articles referenced on SO span 218 FoR across 19 disciplines. Table~\ref{tab:field-table} presents key characteristics of the ten largest fields in terms of SO reference counts. Notably, \textit{Artificial Intelligence} (AI) makes up over 30\% of all references. These AI articles are impactful and recent, averaging 1331 academic citations and an age of 3.75 years at the time of mention. In contrast, SO references from fields such as \textit{Programming Languages} and \textit{Algorithms} tend to be older on average. Such variations indicate the diverse patterns of interacting with academic research on SO, from embracing cutting-edge developments to relying on more established and foundational works. 

% This suggests that SO communities are actively engaging with influential AI research and are abreast with its latest advancements.

% with an average of 1331 academic citations and on average referred to within 3.75 years of publication.

\begin{table}[htbp]
\vspace{-5pt}
\centering
\begin{minipage}[t]{0.60\linewidth} % Adjust the minipage width as necessary
\captionsetup{font=small}
\captionof{table}{Top 10 Fields of Research (FoR)}
\label{tab:field-table}
\vspace{-10pt}
\small
\resizebox{\columnwidth}{!}{%
\renewcommand{\arraystretch}{1.17}
\begin{tabular}{@{}>{\raggedright\arraybackslash}m{0.48\linewidth}m{0.08\linewidth}>{\raggedright\arraybackslash}m{0.18\linewidth}>{\raggedright\arraybackslash}m{0.07\linewidth}>{\raggedright\arraybackslash}m{0.16\linewidth}@{}}
\toprule
Field of Research & \#Ref & \multicolumn{2}{c}{Reference Age $\clubsuit$} & \multirow{2}{*}{\parbox{0.16\linewidth}{\small{Avg.}\\\small{Citation~$\spadesuit$}}} \\ \cmidrule(lr){3-4}
&& \multicolumn{1}{m{0.17\linewidth}}{\footnotesize{Median (Q1, Q3)}} & \multicolumn{1}{c}{\footnotesize{Mean}} & \\ \midrule
\textsf{\small Artificial Intelligence} & 5082 & \multicolumn{1}{c}{\textit{\color[HTML]{3531FF} \textbf{2}} (1, 5)} & \multicolumn{1}{c}{\textit{\color[HTML]{3531FF} \textbf{3.75}}} & 1331.5 \\
\rowcolor[HTML]{EFEFEF} 
\textsf{\small Algorithms (ALG)} & 1373 & \multicolumn{1}{c}{6 (2, 14)} & \multicolumn{1}{c}{9.80} & 428.5 \\
\textsf{\small Prog. Lang. (PL)} & 721 & \multicolumn{1}{c}{4 (2, 12)} & \multicolumn{1}{c}{8.36} & 210.4 \\
\rowcolor[HTML]{EFEFEF} 
\textsf{\small Natural Language Processing (NLP)} & 536 & \multicolumn{1}{c}{\textit{\color[HTML]{3531FF} \textbf{2}} (1, 5)} & \multicolumn{1}{c}{4.59} & 1087.1 \\
\textsf{\small Statistics (Stat)} & 481 & \multicolumn{1}{c}{7 (3, 14)} & \multicolumn{1}{c}{10.36} & 780.2 \\
\rowcolor[HTML]{EFEFEF} 
\textsf{\small Combinatorics} & 470 & \multicolumn{1}{c}{\textit{\color[HTML]{FE0000} \textbf{12}} (4, 23)} & \multicolumn{1}{c}{\textit{\color[HTML]{FE0000} \textbf{15.12}}} & \textit{\color[HTML]{3531FF} \textbf{173.2}} \\
\textsf{\small Parallel Computing} & 443 & \multicolumn{1}{c}{3 (1, 8)} & \multicolumn{1}{c}{6.07} & 219.8 \\
\rowcolor[HTML]{EFEFEF} 
\textsf{\small Mathematical Optimization} & 438 & \multicolumn{1}{c}{6 (2, 12)} & \multicolumn{1}{c}{8.73} & \textit{\color[HTML]{FE0000} \textbf{1356.2}} \\
\textsf{\small Theoretical Computer Science} & 398 & \multicolumn{1}{c}{4 (2, 11)} & \multicolumn{1}{c}{7.51} & 592.4 \\
\rowcolor[HTML]{EFEFEF} 
\textsf{\small Data Mining} & 387 & \multicolumn{1}{c}{4 (2, 8)} & \multicolumn{1}{c}{5.83} & 1104.2 \\
\end{tabular}%
}
\end{minipage}
\hfill
\begin{minipage}[t]{0.34\linewidth} % Adjust the minipage width as necessary
\small
\captionsetup{font=small}
\captionof{table}{Top 15 Venues}
\label{tab:venue-table}
\vspace{-10pt}
\resizebox{\columnwidth}{!}{%
\renewcommand{\arraystretch}{1.08}
\begin{tabular}{@{}m{0.32\linewidth}m{0.16\linewidth}>{\raggedright\arraybackslash}m{0.16\linewidth}m{0.32\linewidth}@{}}
\toprule
Venue & \#Ref & h5i~$\star$ & Field \\ \midrule
\textsf{arXiv*} & 1655 & \textsf{N/A} & \textsf{Various} \\
\rowcolor[HTML]{EFEFEF} 
\textsf{NeurlPS} & 597 & 309 & \textsf{ML} \\
\textsf{ICLR} & 530 & 303 & \textsf{ML} \\
\rowcolor[HTML]{EFEFEF} 
\textsf{CVPR} & 517 & \textit{\color[HTML]{FE0000} \textbf{422}} & \textsf{CV} \\
\textsf{ICML} & 361 & 254 & \textsf{ML} \\
\rowcolor[HTML]{EFEFEF} 
\textsf{ICCV} & 189 & 228 & \textsf{CV} \\
\textsf{ACL} & 174 & 192 & \textsf{NLP} \\
\rowcolor[HTML]{EFEFEF} 
\textsf{EMNLP} & 170 & 176 & \textsf{NLP} \\
\textsf{TPAMI} & 144 & 179 & \textsf{CV} \\
\rowcolor[HTML]{EFEFEF} 
\textsf{ECCV} & 136 & 238 & \textsf{CV} \\
\textsf{JMLR} & 129 & 106 & \textsf{ML} \\
\rowcolor[HTML]{EFEFEF} 
\textsf{NAACL} & 113 & 133 & \textsf{NLP} \\
\textsf{ATC} & 109 & \textit{\color[HTML]{3531FF} \textbf{59}} & \textsf{OS} \\
\rowcolor[HTML]{EFEFEF} 
\textsf{Crypto} & 106 & \textit{\color[HTML]{3531FF} \textbf{59}} & \textsf{Security} \\
\textsf{VLDB} & 95 & 79 & \textsf{DB} \\
\rowcolor[HTML]{EFEFEF}
\textsf{AAAI} & 95 & 212 & \textsf{AI} 
% \textsf{\scriptsize AAAI} & 92 & 212 & \textsf{\scriptsize AI} \\
% \textsf{\scriptsize AISTAT} & 90 & 91 & \textsf{\scriptsize AI} \\
% \rowcolor[HTML]{EFEFEF} 
% \textsf{\scriptsize PLOS One} & 87 & 212 & \textsf{\scriptsize Various} \\
% \textsf{\scriptsize MICCAI} & 81 & 89 & \textsf{\scriptsize CV} \\
% \rowcolor[HTML]{EFEFEF} 
% \textsf{\scriptsize Nature} & 75 & \textit{\color[HTML]{FE0000} \textbf{467}} & \textsf{\scriptsize Various} \\ 
\end{tabular}%

}
\end{minipage} \\
\vspace{2.5pt}
\footnotesize{$\clubsuit$: the duration between publication and the time of every reference \hspace{0.2em} $\spadesuit$: the average academic citation counts of referenced articles \hspace{0.2em} $\star$: h5-index~\cite{h5index} retrieved from Google Scholar; all venues ranked within top 20 of their respective fields, except arXiv}
\vspace{-6pt}
\end{table}

\begin{figure*}[h]
    \centering
    \includegraphics[width=\textwidth]{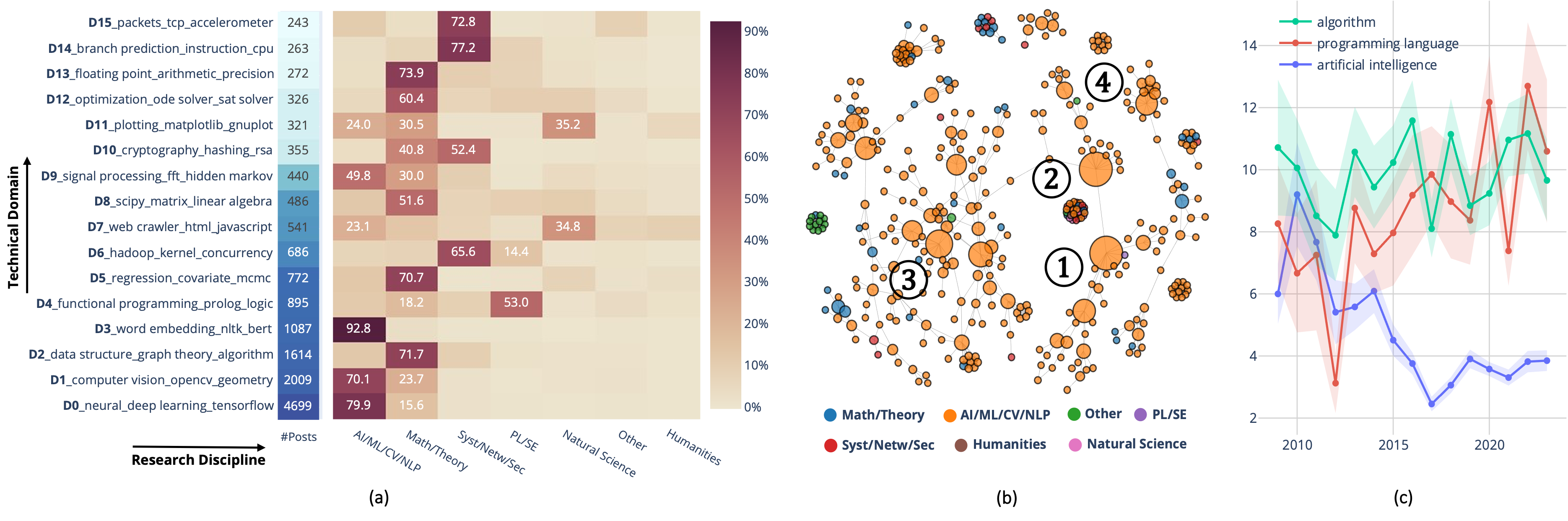}
    \captionsetup{font=small}
    \vspace{-21pt}
    \caption{(a) Heatmap depicting the relationships between technical domains (Y-axis) and research disciplines (X-axis). The blue bar along Y-axis shows the number of posts within each technical domain. Cell \textit{(x, y)} denotes the percentage of papers referenced by domain \textit{x} that originate from discipline \textit{y}. (b) Ten largest components in the co-citation network of academic references on SO. \circled{1}-\circled{4} are key nodes with the highest PageRank score. (c) The average "First-cite Interval" (Y-axis) for articles in three major fields (FoR) cited in SO posts each year (X-axis).}
    \label{fig:grouped}
    % \begin{minipage}{0.46\textwidth}
    %     \includegraphics[width=\linewidth]{img/topic_field_percent.eps}
    %     \vspace{-18pt}
    %     \captionsetup{font=small}
    %     \caption{heatmap}
    %     \label{fig:heatmap}
    % \end{minipage}
    % \hfill
    % \begin{minipage}{0.32\textwidth}
    %     \includegraphics[width=\linewidth]{img/co_citation.eps}
    %     \captionsetup{font=small}
    %     \vspace{-23pt}
    %     \caption{Ten largest components in the co-citation network of academic references on Stack Overflow}
    %     \label{fig:co-citation}
    % \end{minipage}
    % \begin{minipage}{0.2\textwidth}
    %     \includegraphics[width=\linewidth]{img/first_cite.eps}
    %     \captionsetup{font=small}
    %     \caption{}
    %     \label{fig:first-cite}
    % \end{minipage}
    \vspace{-12pt}
\end{figure*}

Table~\ref{tab:venue-table} lists the 15 venues with the most SO references. Consistent with existing studies on Wikipedia citations~\cite{mapofsciencewikipedia}, these venues are highly regarded in their respective fields ($\star$), with an interesting exception of \textit{arXiv*}. We manually inspected a random sample of the referenced articles from \textit{arXiv} and found that many of them were later published in peer-reviewed venues with slightly different titles. An explanation can be that Stack Overflow users tend to (\romannum{1}) integrate academic insights at a fast pace, or (\romannum{2}) prefer open-access content. We observe a similar dominance of AI-related research (e.g., ML, NLP, CV) among the venues, with 12 out of the 15 venues and 51\% of the articles on \textit{arXiv} being related to AI.

% Table~\ref{tab:venue-table} lists the 20 venues with highest SO references. Consistent with existing studies on references in Wikipedia~\cite{mapofsciencewikipedia}, these venues are highly regarded in their respective fields, with an interesting exception of \textit{arXiv}. We observe a similar trend of AI dominance among the most referred venues (14 out of the 20 venues) along with 51\% of the articles on \textit{arXiv} being related to AI. We manually inspected the cited papers from \textit{arXiv} and found that many of them were later published in peer-reviewed venues with slightly different titles. An explanation for this could be that Stack Overflow users (\romannum{1}) integrate academic insights at a fast pace, and (\romannum{2}) prefer open-access content. 

\vspace{5pt}
\noindent
\emph{\textbf{RQ 2}: Which parts of Stack Overflow rely on academic sources?} 

% how users interested in different technical domains were associated with posts citing academic literature.
% maybe no transition
We discerned which technical domains and user communities were associated with the posts citing academic literature on Stack Overflow by analyzing the overarching themes of these discussions. Although using the user-generated tags attached to each SO post for topic modeling may seem intuitive, prior research suggests that such tags often fall short in reflecting the actual discussions accurately~\cite{Wang2014EnTagRecAE}. Moreover, SO has over 65,000 existing tags that are too parochial to capture each post's broader themes and areas of interest. Following existing works~\cite{sobertopic}, we utilized BERTopic~\cite{Grootendorst2022BERTopicNT} to categorize the discussions into coarser technical domains. Initially, the model was fine-tuned for better granularity and coherence, producing 109 preliminary topic clusters. Two authors then discussed and qualitatively merged related topics into broader \textit{domains}. For example, clusters about \textit{Named Entity Recognition} and \textit{Sentiment Analysis} were grouped under the domain of \textit{NLP} (D3). Eventually, we compiled all SO topics into 16 technical domains that serve as focal points of interest for distinct communities on SO.

The leftmost column in Figure~\ref{fig:grouped}(a) shows the number of posts containing academic references within each technical domain (\textit{Y-axis}). Notably, the domains of D0 (\textit{machine learning}), D1 (\textit{vision/ graphics}), and D2 (\textit{algorithms}) were most actively incorporating academic knowledge. Whereas more traditional and application-oriented domains such as D15 (\textit{data communications}) and D14 (\textit{computer architecture}) exhibit less integration of academic research.

% Notably, discussions around \textit{machine learning} (D0, D3), \textit{vision/graphics} (D1), and \textit{algorithms} (D2) are most actively incorporating scholarly knowledge. Whereas more application-oriented domains such as \textit{data communications} (D15) and \textit{computer architecture} (D14) exhibit less reliance on academic research.

% which is named by the three most representative keywords identified by c-TF-IDF to succinctly summarize its core topics

% Moreover, a number of references from these domains are actually IEEE Standard Specifications, which, despite having DOIs, are not generally considered as academic literature.

% # s/o posts in total that cite papers (12k), 15k references
% the topics - briefly talk about whats in each topic
% blue bar: distribution of posts across topics that cite papers
% ? # papers cited per topic

\vspace{3pt}
\noindent
\emph{\textbf{RQ 3}: How do various Stack Overflow communities interact with academic research?}

To examine how different SO communities interact with academic research, we mapped the citation flow between 218 research fields (FoR) and the 16 technical domains. However, the resulting bipartite network is hard to interpret and visualize due to its high dimensionality. We simplified its complexity by aggregating the 218 research fields into seven broader \textit{Disciplines}. The simplification was carefully executed to preserve the underlying citation patterns, guided by two criteria: (\romannum{1}) the similarity in how different FoRs are referenced together across technical domains, signaled by the pairwise Spearman's correlation~\cite{spearman} (where high correlations suggest similar citation patterns), and (\romannum{2}) their hierarchical relationships within an FoR ontology~\cite{Salatino2018TheCS}, as to ensure the aggregation also respects established disciplinary structures. For example, \textit{Computer Network}, \textit{Distributed Computing}, and \textit{Comp. Security} were grouped into the \textit{Syst/Netw/Sec} category for their high inter-correlations in citation patterns (above 0.6) and shared academic lineage.

% To explore how different Stack Overflow communities interact with academic research, we mapped 218 research fields (FoR) to the 16 technical domains citing them. However, the resulting bipartite network is hard to interpret and visualize due to its high dimensionality. We simplified its complexity by aggregating the 218 research fields into seven broader \textit{Disciplines}. The simplification was carefully carried out to preserve the underlying citation patterns, guided by two criteria: (\romannum{1}) the similarity in how different FoRs are cited together across technical domains, signaled by the pairwise Spearman's correlation (where high correlations suggest similar citation patterns), and (\romannum{2}) their hierarchical relationships within an FoR ontology~\cite{Salatino2018TheCS}, as to ensure the aggregation also respects established disciplinary structures. For example, \textit{Computer Network}, \textit{Distributed Computing}, and \textit{Computer Security} were grouped into the \textit{Syst/Netw/Sec} category for their high inter-correlations of citation patterns (above 0.6) and shared academic lineage.

Figure~\ref{fig:grouped}(a) illustrates the relationships between 16 technical domains and seven research disciplines. Cell $(x, y)$ denotes the percentage of papers referenced by domain $x$ that originate from discipline $y$. For example, 40.8\% of the articles referenced in domain D10 are from the \textit{Math/Theory} discipline. This figure reveals that SO discussions typically rely on one single research discipline, with notable exceptions being D7 (\textit{web scraping}) and D11 (\textit{data visualization}), where the distributions are relatively even. We analyzed sample posts from these two domains and discovered that posts within D7 often seek guidance on downloading research articles or scraping bibliographic data programmatically, while the discussions in D11 typically revolve around replicating data visualizations in scientific articles. These activities (downloading and visualizing) are universal and not confined to any field, leading to the balanced distribution observed. In such cases, academic references primarily serve as illustrative examples or supplementary materials, rather than as integrated sources of knowledge for problem-solving.

We further explored which articles were referenced together on SO and mapped the structure of scientific knowledge through a co-citation analysis. Each node in the co-citation network represents an academic article and is connected to another if they were jointly referenced by the same SO post. The resulting network was highly fragmented and sparse, with 2541 nodes (23.7\% of all cited articles on SO) connected by merely 3089 edges. This limited connectivity likely stems from the highly specialized nature of many SO discussions~\cite{nicheso}, which often demand niche and domain-specific knowledge, resulting in less overlap among the cited articles. Confirming this hypothesis, however, requires a more detailed contextual analysis of the isolated dyads and fragments within the network, a task we reserved for future studies. 

Meanwhile, our current study focused on the ten largest components in the co-citation network, comprising 381 nodes and 1358 edges, as depicted in Figure~\ref{fig:grouped}(b). The size of a node reflects the article's PageRank score, denoting its importance and influence within the network. Notably, we observed that pivotal and pioneering works---those that lay new foundations and advance the field---are associated with highest PageRank scores. Noteworthy examples include papers that introduced the \circled{1} \textit{Transformer Architecture} \href{https://arxiv.org/abs/1706.03762}{\scriptsize{\faExternalLink}}, \circled{2} \textit{BERT language model} \href{https://aclanthology.org/N19-1423/}{\scriptsize{\faExternalLink}}, \circled{3} \textit{Deep Residual Network} \href{https://arxiv.org/abs/1512.03385}{\scriptsize{\faExternalLink}}, and \circled{4} \textit{Generative Adversarial Network} \href{https://arxiv.org/abs/1406.2661}{\scriptsize{\faExternalLink}}. Our qualitative analysis found that users often cite these articles alongside others in a post to provide essential background knowledge for understanding its content. A practice that makes academic findings more accessible to a wider audience, indicating SO's role in bridging the knowledge gap between forefront academic research and software practitioners.

% heatmap - relation between topics and field
% timeseries - how quickly s/o cites published papers

% =========== 
% network graph
% ===========

\vspace{3pt}
\noindent
\emph{\textbf{RQ 4}: How quickly does Stack Overflow integrate academic research?}

We analyzed the pace at which academic articles were referenced on Stack Overflow by tracking the "First-cite Interval" --- the time lag between a paper's first mention on SO and its publication date. This backward tracing approach circumvents the potential right censoring bias~\cite{ivorytower}. On average, it takes an article 6.6 years to be recognized on SO, although this interval varied significantly across research fields. For instance, AI-related papers were typically referenced within 3.7 years, whilst PL papers have a longer latency of 9.3 years. Figure~\ref{fig:grouped}(c) further illustrates the trend of diffusion rates of three major research fields (see Table~\ref{tab:field-table}) over time. Prior to 2010, these fields exhibited similar first-cite intervals of around 6-10 years (Note that SO was launched in late 2008). However, from 2010 to 2017, the rate at which AI articles were referenced accelerated rapidly. This timeframe coincides with significant breakthroughs in AI, such as the \textit{Adam optimizer}, \textit{Batch normalization}, \textit{Attention mechanism}, etc. Conversely, the First-cite Interval for Algorithms papers saw little changes, while that for PL papers even noticeably increased, implying that earlier foundational works in these areas might hold more relevance in Stack Overflow discussions.

% The distributions of first-cite intervals between AI and PL are statistically different, with a $p$-value $< 10^{-5}$ from the Mann-Whitney U test [$U=37137, z=-6.49$].

\section{Discussion and Conclusion}
This study presents the first large-scale analysis of academic references on Stack Overflow, aiming to understand how scholarly knowledge appears in this practitioner-centric community. In this section, we discuss the implications of our results.

\vspace{2pt}
\noindent
\textbf{Divergent trajectories.}\hspace{0.4em} Our analysis suggested that the research trajectories in certain fields do not always align with the practical discussions on Stack Overflow. For instance, discussions in D15 (\textit{data communications}) referred mostly to \textit{computer network} articles (as expected), but the number of references is low, and there's a noticeable preference for older publications. This hints at a potential disconnect from the latest developments in this field. Additionally, a significant number of the articles referenced in domains such as D13 (\textit{floating point}), D14 (\textit{computer architecture}), and D15 were in fact technical documents (with DOI) like IEEE Standard Specifications and IETF Request for Comments (RFCs). These documents are meant for establishing technical foundations and addressing engineering challenges~\cite{rfc}, rather than presenting new scientific findings. Considering that systems and networking are relatively mature fields with close ties to the industry, it is reasonable that professionals in these domains find greater value in well-tested, experiential knowledge over yet-to-be-proven academic innovations.

% Although these documents have DOIs, they are typically not classified as academic literature. This may suggest that communities with greater focus on more traditional and engineering domain find more value in industry standards and experiential knowledge over academic research.

% traditional application and engineering challenges find more value in industry standards and experiential knowledge over academic research, which further points to a possible divergence between cutting-edge academic findings and the practical, day-to-day challenges faced by professionals in certain technical domains.

\vspace{3pt}
\noindent
\textbf{Feasibility as an Altmetric.}\hspace{0.4em} Academic references on Stack Overflow offer a novel lens to observe how scholarly knowledge diffuses into the developer communities, suggesting their potential as an alternative metric (altmetric) for gauging the industry impact and practicality of academic research. 
% Our findings provide preliminary support for this vision.

We observed a tangible link between the volume of SO references and the impact of scholarly contributions in real-world settings. For example, articles that emerged as the central nodes in the SO co-citation network are typically groundbreaking seminal works with extensive application in the industry. Furthermore, the time interval between an article's publication and its acknowledgment on SO is comparatively shorter than that observed with other altmetrics, such as Patent~\cite{ivorytower} and Wikipedia~\cite{mapofsciencewikipedia} citations, suggesting that SO references may offer a more immediate measure of the impact.

However, there are challenges to consider. For instance, not all references signify an effective transfer of scholarly knowledge, as seen in D11 (\textit{data visualization}) and D7 (\textit{web scraping}), where academic references serve merely as contextual support. As such, to make SO a meaningful indicator of research impact in the industry, future research needs to develop rigorous methods to evaluate the intention and contribution of these academic references. Specifically, to contextually assess whether a SO reference genuinely facilitates the diffusion of knowledge and provides tangible solutions to the challenges faced by software practitioners and developers.

% Comparing with other Altmetrics like Wikipedia citations, Stack Overflow references are more timely and up to date, while prior research found Wikipedia to be a "slow altmetric indicator" and 

% As a platform primarily used by software developers to discuss practical challenges, Stack Overflow has been widely used as a proxy for studying developers' behavior. 

% As a hub for discussing practical challenges in the software development industry, Stack Overflow surprisingly hosts plentiful discussions that refer to academic research. In this work, we performed the first large-scale analysis of these academic references, 

% aiming to understand how scholarly knowledge diffuses into this developer-centric community and contributes to solving practical, real-world problems. Our findings reveal diverse patterns of academic integration across different technical domains and user communities. This investigation serves as an initial step towards exploring the feasibility of using Stack Overflow references as an Altmetric for evaluating the industrial impact of a scholarly work.

\bibliographystyle{ACM-Reference-Format}
\bibliography{sample-base}

\end{document}